\documentclass[sigplan,screen,nonacm]{acmart}

\AtBeginDocument{%
  }

\usepackage{booktabs}
\usepackage{filecontents}
\usepackage{graphicx}
\usepackage[many]{tcolorbox}

\usepackage{mathtools,amssymb,latexsym,amsfonts,stmaryrd}
\usepackage{bbm}
\usepackage{pbox}
\usepackage{xfrac}
\usepackage{booktabs}
\usepackage{xcolor}
\usepackage{threeparttable}
\usepackage{amsthm}
\usepackage{float}
\usepackage{multirow}
\usepackage{tikz}
\usepackage{mathrsfs}
\usepackage{mathpartir}
\usepackage{subcaption}
\usepackage{algorithm}
\usepackage{algorithmicx}
\usepackage{url}
\usepackage{syntax}
\usepackage{framed}
\usepackage[noend]{algpseudocode}
\usepackage{flushend}
\usepackage{listings}
\usepackage{moresize}
\usepackage{wrapfig}
 
\usepackage{seqsplit}
\usepackage{alltt}
\usepackage{xspace}
\usepackage{enumitem}

\newcommand{\revise}[2]{{\color{red}{\ifx&#1&\else- #1\fi}} {\color{ForestGreen}{\ifx&#2&\else+ #2\fi}}}%
\renewcommand{\revise}[2]{#2}%

\usetikzlibrary{arrows,patterns, decorations.pathreplacing}

\newcommand{\ignore}[1]{}

\usepackage{pifont}

\usepackage{amsmath, listings, amsthm, amssymb, proof, xspace}
\usepackage{bbding}

\title{Position Paper: Programming Language Techniques for Bridging LLM Code Generation Semantic Gaps}

\author{Yalong Du}
\email{23S154041@stu.hit.edu.cn}
\affiliation{%
  \institution{Harbin Institute of Technology, Shenzhen}
  \city{Shenzhen}
  \country{China}
}

\author{Chaozheng Wang}
\authornote{Corresponding author.}
\email{adf111178@gmail.com}
\affiliation{%
  \institution{The Chinese University of Hong Kong}
  \city{Hong Kong}
  \country{China}
}

\author{Huaijin Wang}
\email{hwangdz@cse.ust.hk}
\affiliation{%
  \institution{The Hong Kong University of Science and Technology}
  \city{Hong Kong}
  \country{China}
}

\begin{document}

\begin{abstract}
Large Language Models have demonstrated remarkable capabilities in automated code generation, yet their statistical nature and black-box characteristics create significant semantic gaps manifested through syntax errors, semantic hallucinations, and reliability concerns. This position paper argues that principled integration of Programming Language (PL) techniques is essential for bridging these gaps. Through structured program representations, formal correctness guarantees, and robust verification mechanisms, PL techniques can elevate LLM-generated code from statistical pattern matching to truly reliable and trustworthy levels. This integration is crucial for developing systems that generate code that is not only functionally correct but also interpretable, verifiable, and ultimately trustworthy.
\end{abstract}

\maketitle

\section{Introduction}

The emergence of Large Language Models (LLMs) has catalyzed a paradigm shift in automated code generation. These models exhibit remarkable proficiency in comprehending and synthesizing diverse programming constructs, offering transformative solutions that substantially enhance developer productivity while addressing the escalating demands of software development~\cite{bui2025correctness, wong2025decllm, wang2023prompt,li2025differentiation, wang2024systematic,ji2024testing,li2024empirical}. Contemporary development ecosystems increasingly incorporate LLMs across a spectrum of tasks, from code completion~\cite{li2022cctest, wang2025beyond, wang2025rag, li2025api,wong2023refining} and program repair~\cite{jelodar2025large} to sophisticated code synthesis applications~\cite{ashrafi2025enhancing}.

Nevertheless, despite their impressive generative prowess, empirical evidence reveals a troubling phenomenon: LLM-generated code frequently exhibits a higher prevalence of errors and security vulnerabilities compared to human-authored code. These deficiencies manifest across multiple dimensions, encompassing fundamental syntax errors, hallucinated non-existent functions, and intricate logical inconsistencies~\cite{huynh2025large}. The fundamental ``black-box'' architecture of LLMs precludes intrinsic guarantees regarding output correctness or reliability~\cite{cramer2025verifying}. This opacity presents formidable obstacles for deployment in production environments, particularly within domains where reliability, security, and auditability constitute critical requirements.

This position paper contends that systematic integration of Programming Language (PL) techniques represents an essential pathway for addressing the semantic gaps inherent in LLM-based code generation. Through the strategic application of structured program representations (including intermediate representations, abstract syntax trees, and control/data flow graphs), formal correctness frameworks (encompassing formal methods and type systems), and comprehensive verification methodologies (such as advanced testing and static analysis), PL techniques can fundamentally transform LLM-generated code from probabilistic pattern matching to demonstrably reliable and verifiable software artifacts.\looseness=-1

\section{Semantic Gaps in LLM Code Generation}

\subsection{Syntactic and Semantic Errors}

LLMs frequently generate code that fails to compile or contains syntax errors, including incorrect indentation, improper conditional or loop structures, and assignment errors. Beyond mere syntactic issues, deeper semantic errors are also prevalent. These include API misuse (e.g., missing imports, incorrect or redundant parameters) and invalid reference errors, where generated code attempts to access undefined variables or non-existent object members. These problems arise because LLMs, as statistical models, may produce code that appears syntactically plausible but is functionally incorrect or logically unsound~\cite{bui2025correctness}.

Code hallucination is a particularly challenging issue where LLMs generate incorrect, nonsensical, or unprovable information that is difficult for human users to identify and correct~\cite{su2025code}. A recent taxonomy categorizes these hallucinations into syntactic hallucinations, runtime execution hallucinations, functional correctness hallucinations, and code quality hallucinations~\cite{lee2025hallucination}. For example, functional correctness hallucinations include defects in algorithm implementation (e.g., missing edge cases, incorrect arithmetic operations) and non-compliance with specified requirements~\cite{wang2025towards}. Code quality hallucinations involve issues such as improper resource handling or the introduction of security vulnerabilities.

Errors in LLM-generated code exhibit a clear hierarchical structure. It begins with easily detectable syntax errors, progresses to more subtle runtime execution hallucinations~\cite{lee2025hallucination}, and culminates in the most challenging functional correctness hallucinations~\cite{wang2025towards} and code quality hallucinations. This progression suggests that LLMs excel at learning surface-level patterns but lack genuine deep understanding of computational logic and program semantics. Their statistical nature enables them to generate plausible-looking code structures that, upon execution or deeper analysis, prove fundamentally flawed. This multi-layered error characteristic demands a multifaceted validation approach. Simple syntax errors can be caught by compilers/linters, but deeper semantic and functional errors require sophisticated program analysis techniques capable of reasoning about program behavior, not just its form. This further reinforces that programming language techniques are crucial for comprehensive quality assurance of LLM-generated code.

\subsection{Lack of Deep Code Understanding}

Notwithstanding substantial technological progress, LLMs fundamentally process code as sequential token streams, thereby disregarding its intrinsic structural properties and failing to achieve profound semantic comprehension~\cite{learningnova}. Empirical investigations reveal that while LLMs demonstrate competency in basic syntactic recognition (analogous to abstract syntax tree parsing capabilities), their capacity for understanding static program behavior—and particularly dynamic execution semantics—remains severely constrained. These models exhibit tendencies toward hallucination when interpreting semantic code structures, frequently relying on heuristic approximations rather than precise instruction-level logical reasoning~\cite{ma2023lms}. This limitation becomes particularly pronounced in scenarios demanding sophisticated control flow analysis, loop invariant maintenance, and accurate tracking of variable state evolution and transformations~\cite{jiang2025can}.

\subsection{Reliability and Security Concerns}

The prevalence of defects and security vulnerabilities in LLM-generated code constitutes a critical systemic issue~\cite{bui2025correctness}. Empirical studies demonstrate that systems such as GitHub Copilot produce approximately 40\% of code containing potential vulnerabilities~\cite{yang2024robustness}. This statistic presents substantial risks for deploying LLM-generated artifacts within mission-critical production environments. Additionally, safeguarding the intellectual property embedded within commercial LLMs and their generated outputs presents formidable challenges, with documented risks of model extraction and imitation attacks necessitating robust watermarking and verification technologies~\cite{li2023protecting}.

The characterization of LLMs as black box~\cite{cramer2025verifying} exhibits direct correlation with observed patterns of unreliability~\cite{lu2025safe}, security vulnerabilities~\cite{yang2024robustness,lu2025less}, and intellectual property concerns~\cite{li2023protecting,LuLHOW024}. When code generation processes remain opaque, determining the root causes of errors or vulnerabilities, or establishing proper attribution for generated code, becomes virtually intractable. This transparency deficit fundamentally undermines trust, impedes effective debugging practices, and complicates the legal and ethical frameworks surrounding LLM outputs~\cite{wang2024navrepair}. These considerations underscore that achieving explainability and verifiability transcends academic interest to become a practical imperative for responsible, widespread adoption of LLM-generated code in critical applications. Viable solutions must either illuminate the black box through techniques such as internal state analysis or provide robust external guarantees regarding output properties—a domain where programming language techniques demonstrate particular strength.

\section{Structured Program Representations}

\subsection{Leveraging Intermediate Representations (IRs)}

Intermediate Representations, fundamental to compilers, significantly enhance neural network program embeddings for downstream tasks like clone detection and program repair~\cite{li2022cctest, jiang2025can}. IRs function as a semantic normalization layer; by compiling source code, syntactic variations such as variable names and coding styles are minimized, making functionally equivalent code appear more similar~\cite{li2022unleashing}. This normalization provides LLMs with cleaner, more semantically focused data, enabling them to concentrate on underlying logic rather than superficial syntax. This highlights the value of a multimodal input strategy, where LLMs learn from both high-level source code and low-level, normalized IRs to gain a more robust and generalizable understanding of programs.

Despite these benefits, a critical semantic gap remains, as LLMs struggle with deep, instruction-level reasoning about IRs, particularly concerning precise control flow, loop invariants, and dynamic execution behavior~\cite{bui2025correctness, jiang2025can,li2025reasoning}. While LLMs can parse IR syntax, they cannot yet perform the rigorous logical deductions that traditional compilers do, indicating they represent IRs without fully grasping their formal semantics. Therefore, simply using IRs as an input is insufficient. Future research must focus on designing LLM architectures or training objectives, such as neurosymbolic methods or pre-training tasks that simulate IR execution, to explicitly teach formal reasoning. This will be crucial for moving LLMs beyond statistical pattern matching toward computationally sound, logical deduction.

\subsection{Structural and Semantic Program Representations}

Contemporary LLMs predominantly process code as linear token sequences, consequently neglecting its fundamental structural characteristics. The AST-T5 framework introduces an innovative pre-training methodology that explicitly exploits Abstract Syntax Trees (ASTs) to augment code generation, transpilation, and comprehension capabilities~\cite{gong2024ast,DBLP:journals/tosem/GaoGHZNXL23,wang2023reef,sem2vec2023}. This ``structure-aware pre-training'' paradigm implements AST-aware segmentation and span corruption objectives to systematically preserve and reconstruct code structure, yielding substantial performance enhancements across code-to-code transformation tasks.

Empirical investigations demonstrate that although LLMs exhibit competence in syntactic recognition (analogous to AST parsing mechanisms), they encounter substantial difficulties in comprehending deeper code semantics, particularly dynamic semantic properties~\cite{wang2024sanitizing,wang2024llmdfa}. These models demonstrate susceptibility to hallucinations when processing complex semantic code structures~\cite{ma2023lms}. The explicit integration of Control Flow Graphs (CFGs) and Data Flow Graphs (DFGs) enables the capture of critical execution pathways and data dependencies, which constitute essential components for sophisticated semantic understanding, error detection, and diverse program analysis applications~\cite{wang2022enriching}. The augmentation of LLM prompts with semantic facts derived from these structured representations has demonstrated measurable improvements in tasks like code summarization~\cite{ahmed2024automatic,gao2023makes}.\looseness=-1

These investigations illuminate a distinct evolutionary trajectory in LLM code comprehension: progressing from elementary token sequence processing (the foundational approach) to Abstract Syntax Tree (AST) integration for hierarchical structural understanding~\cite{gong2024ast}, and subsequently advancing to Control Flow Graph (CFG) and Data Flow Graph (DFG) incorporation for behavioral and dependency modeling~\cite{ma2023lms}. This progression directly parallels the abstraction hierarchies employed in conventional compiler design and program analysis frameworks. The recognition that raw token processing proves inadequate for profound semantic understanding mandates explicit encoding and manipulation of these enriched, graph-based program representations. Future LLM architectures and pre-training methodologies should advance toward native integration of these graph structures through sophisticated Graph Neural Networks (GNNs) or alternative mechanisms specifically designed for processing and reasoning about graph-structured data~\cite{sem2vec2023,yu2020order,binaug2024}, rather than relying exclusively on sequential token manipulation. The ultimate objective involves endowing LLMs with ``compiler-analogous'' comprehension of code's underlying logic and behavioral characteristics.

Despite advances in static structural understanding through ASTs, CFGs, and DFGs, LLMs continue to encounter substantial challenges with dynamic semantics and accurate runtime behavior prediction. This phenomenon indicates fundamental constraints in program execution simulation, temporal state change tracking, and complex interaction handling that emerge exclusively during runtime. Their reasoning processes frequently remain anchored to static features, resulting in ``hallucinations and fabrication of non-existent facts'' when interpreting semantic code structures. Addressing this limitation necessitates the integration of dynamic analysis techniques and runtime feedback mechanisms into LLMs' learning and generation cycles. This integration may encompass symbolic execution~\cite{li2025large} or alternative dynamic program analysis methodologies to provide concrete execution traces and counterexamples, thereby grounding LLMs' abstract understanding in actual program behavior and enhancing their capacity for dynamic property reasoning~\cite{li2024accuracy}.

\section{Formal Methods and Type Systems}

\subsection{Formal Verification for Provable Correctness}

Formal verification techniques provide mathematical rigor for ensuring the reliability of LLM-generated code. Tools such as Marmaragan demonstrate the feasibility of leveraging LLMs to generate SPARK annotations for programs, enabling formal verification with approximately 50.7\% success rate~\cite{cramer2025verifying}. Proof assistants like Lean, Coq, and Isabelle can verify LLM reasoning and provide automated feedback, addressing challenges in advanced mathematics and the inherent unverifiability of informal LLM outputs~\cite{yang2024formal}. LLMs can generate program invariants alongside test inputs for modular code review and analysis~\cite{sun2025classinvgen}, while contract synthesizers can automatically generate pre/post-condition contracts for multi-language systems~\cite{chen2025polyver}.

The synergy between LLMs and formal methods addresses the fundamental issue of LLM "hallucinations" by providing explicit semantic truths through rigorous mathematical proofs~\cite{lee2025hallucination}. This represents a shift from probabilistic to provable correctness, where LLMs generate code candidates and formal properties that are then rigorously verified by external formal verifiers.

LLMs can overcome the traditional scalability challenges of formal methods by automating the generation of specifications, invariants, and proof steps~\cite{zhong2025approach}. This automation makes formal verification more practical for industrial applications by reducing the manual effort required for specification writing while maintaining critical oversight for essential components.\looseness=-1

\subsection{Type System-Guided Generation}

LLMs frequently generate non-compilable code because their token-based inference does not model formal programming language aspects, particularly type rules. Type-constrained decoding methods address this through novel prefix automata that enforce well-typedness during generation, significantly reducing compilation errors~\cite{mundler2025type}.

For dynamic languages like Python, hybrid approaches such as HiTyper combine static inference with deep learning using Type Dependency Graphs (TDGs) to integrate inference rules and validate neural predictions, outperforming pure deep learning models~\cite{peng2022static}. Integration with Language Server Protocols provides real-time type and binding information for semantic consistency~\cite{blinn2024statically}.

Type systems provide semantic grammar that extends beyond basic syntax, guiding LLMs to generate type-safe and semantically consistent code. This elevation from token prediction to type-aware generation suggests future programming languages may incorporate richer type annotations to provide more fine-grained information to LLMs. The hybrid neuro-symbolic approach, combining traditional programming language techniques with LLM capabilities, demonstrates superior results for complex programming tasks across various language paradigms.

\section{Advanced Program Analysis for Interpretability}

\subsection{Automated Testing and Repair Frameworks}

Automated testing and repair represent critical components for enhancing the quality and reliability of LLM-generated code. The CCTEST framework exemplifies effectiveness in testing and repairing code completion systems within black-box environments, utilizing innovative Program Structure Consistency (PSC) mutations to identify inconsistencies and automatically repair outputs, thereby achieving substantial accuracy improvements~\cite{li2022cctest, lam2025codecrash}.

Contemporary Automated Program Repair (APR) techniques are being strategically applied to LLM-generated code, with particular emphasis on ``last-mile improvements'' to address residual errors. Following error localization, LLMs can predict alternative candidate code snippets, effectively transforming the program repair process into a code generation challenge. The most pragmatic and effective approach for ensuring correctness, considering the inherent statistical characteristics and potential unreliability of LLM outputs, involves implementing a systematic ``generate-verify'' cycle~\cite{gao2024search}.
Fuzzing methodologies, particularly those enhanced through LLM-synthesized input generators such as G2FUZZ, offer innovative and comprehensive approaches for grammar-aware testing of non-textual inputs. These approaches demonstrate improved code coverage and enhanced error discovery rates while substantially reducing LLM computational costs. Recent research indicates that LLMs have evolved beyond mere code generation tools, becoming active and integral participants throughout the entire software testing and repair lifecycle~\cite{zhang2025low, campos2025empirical, li2024split}.

\subsection{Static Analysis Enhancement}

Static analysis serves as a fundamental technique for software vulnerability detection and error diagnosis. Nevertheless, conventional static analysis tools frequently generate excessive false positives due to oversimplified vulnerability modeling and over-approximation of path and data constraints~\cite{DBLP:conf/icse/WenCGZZL23/xincheng}. LLMs demonstrate significant potential in comprehending code semantics and API usage patterns, offering promising opportunities to enhance static analysis through more sophisticated code understanding and optimized analysis constraints~\cite{li2025hitchhiker}.

Advanced frameworks such as BugLens guide LLMs through structured reasoning processes to evaluate security impact and verify constraints within source code, achieving substantial improvements in static analysis precision and reducing false positives, while occasionally discovering previously unreported vulnerabilities. Although traditional static analysis tools possess considerable power, they often encounter scalability challenges and elevated false positive rates due to inherent limitations in reasoning about complex semantic properties. LLMs can function as ``semantic enhancement layers'' to help filter false positives and provide more nuanced understanding of potential vulnerabilities.

\subsection{Causal Analysis and Internal State Analysis}

Comprehensively evaluating and interpreting LLM code generation capabilities presents inherent challenges due to their architectural complexity and black-box characteristics. Causal analysis-based methodologies provide systematic approaches for examining causal relationships between LLM input prompts and generated code, delivering valuable insights into LLM effectiveness while helping end users understand prediction mechanisms~\cite{ji2025causality,gao2023two,DBLP:journals/corr/abs-2404-15596/xincheng}.
Emerging research investigates leveraging LLMs' internal states to assess code correctness and proactively identify defects. The fundamental hypothesis suggests that these internal states contain meaningful signals reflecting the reliability of generated code, enabling earlier detection of potential errors and significantly reducing computational costs and time requirements for quality assurance~\cite{bui2025correctness}. Furthermore, accurate judgment of code correctness can also further assist in attacking and defending code models~\cite{li2023protecting,li2023feasibility}.

Current evaluation methodologies for LLM-generated code predominantly concentrate on final output results. However, achieving genuine interpretability necessitates understanding the rationale behind specific outputs and the mechanisms through which LLMs arrive at particular solutions. Rather than relying exclusively on post-generation verification, analyzing LLMs' internal states during code generation can provide valuable ``in-process signals'' regarding the correctness and reliability of evolving code.

\section{Conclusion}

While Large Language Models have shown remarkable promise in automated code generation, their statistical nature and black-box characteristics fundamentally limit their reliability and trustworthiness in production environments. We argue that the future of LLM-based code generation lies not in incremental improvements to existing approaches, but in a paradigmatic shift toward deep integration of programming language theory with neural architectures. This integration must prioritize semantic understanding, formal verification, and interpretability to bridge the critical gap between statistical pattern matching and reliable software synthesis.

\bibliographystyle{plain}
\bibliography{./bib/code, ./bib/yalong, bib/czwang}

\end{document}